\documentstyle[epsfig,preprint,aps]{revtex}
\begin{document}
\draft
\def\bra#1{{\langle #1{\left| \right.}}}
\def\ket#1{{{\left.\right|} #1\rangle}}
\def\bfgreek#1{ \mbox{\boldmath$#1$}}
\title{Electromagnetic form factors of the bound nucleon}
\author{D.H. Lu, K. Tsushima, A.W. Thomas and A.G. Williams}
\address{Department of Physics and Mathematical Physics\break
 and
        Special Research Centre for the Subatomic Structure of Matter,\break
        University of Adelaide, Australia 5005}
\author{K. Saito}
\address{Physics Division, Tohoku College of Pharmacy,\break
Sendai 981-8558, Japan}
\maketitle

\vspace{-9.7cm}
\hfill ADP-98-39/T311
\vspace{9.7cm}

\begin{abstract}
We calculate electromagnetic form factors of the proton bound in
specified orbits for several closed shell nuclei. The quark structure 
of the nucleon and the shell structure of the finite nuclei are
given by the QMC model. We find that orbital electromagnetic form factors
of the bound nucleon deviate significantly from those of the free nucleon.
\end{abstract}



Whether or not quark degrees of freedom play any significant role beyond
conventional nuclear theory (involving baryons and mesons) is a fundamental
question in strong interaction  physics.
Tremendous efforts have been devoted to the study of medium modifications 
of hadron properties\cite{matter95}.
The idea that nucleons might undergo considerable change of their internal 
structure in a baryon-rich environment has been stimulated by
a number of experiments, e.g., the variation of nucleon  structure 
functions in lepton deep-inelastic scattering off nuclei 
(the nuclear EMC effect)\cite{EMC}, 
the quenching of the axial vector coupling constant, $g_A$, 
in nuclear $\beta$-decay\cite{gA}, and the missing strength 
of the response functions in nuclear quasielastic  electron 
scattering\cite{quasi}. 
Though the conventional interpretation arising  through polarization effects
and other hadronic degrees of freedom ($\Delta$-excitations, meson exchange 
currents, etc.) cannot be  ruled out 
at this stage\cite{Mulders90,MEC}, 
it is rather interesting to explore the possibilities of a
change in the internal structure of the bound nucleon.

There have been several effective Lagrangian approaches in the literature
dealing with modifications of the nucleon size and electromagnetic properties 
in medium\cite{Meissner89,medium}.
All these investigations found that nucleon electromagnetic form factors are 
suppressed and the rms radii of the proton somewhat increased in 
bulk nuclear matter --- in addition to hadron mass reductions. 
In Ref.\cite{medium}, we examined medium modifications of nucleon 
electromagnetic properties in nuclear matter, using the quark-meson
coupling model (QMC)\cite{Guichon,finite}. 
The self-consistent change in the internal structure of a bound nucleon
is consistent with the constraints from $y$-scaling data\cite{sick} 
and the Coulomb sum rule\cite{coulomb}.
In this letter, we calculate 
electromagnetic form factors for a nucleon bound in specific, shell model
orbits of realistic finite nuclei.
This is of direct relevance to quasielastic scattering measurements 
underway at TJNAF\cite{TJNAF}.   

The details for solving QMC  for finite nuclei can be found
in Ref.~\cite{finite}. Here we briefly illustrate the essential features 
of this work.
For the calculation of the nucleon shell model wave functions, 
the QMC model for spherical finite nuclei, in mean-field approximation, 
can be summarized in an effective Lagrangian density\cite{finite}
\begin{eqnarray}
{\cal L}_{QMC}&=& \overline{\psi}({\vec r}) [i \gamma \cdot \partial 
- m_N + g_\sigma(\sigma({\vec r})) \sigma({\vec r})  
- g_\omega \omega({\vec r}) \gamma_0 \nonumber \\
&-& g_\rho \frac{\tau^N_3}{2} b({\vec r}) \gamma_0 
- \frac{e}{2} (1+\tau^N_3) A({\vec r}) \gamma_0 ] \psi({\vec r}) \nonumber \\
&-& \frac{1}{2}[ (\nabla \sigma({\vec r}))^2 + 
m_{\sigma}^2 \sigma({\vec r})^2 ] 
+ \frac{1}{2}[ (\nabla \omega({\vec r}))^2 + m_{\omega}^2 
\omega({\vec r})^2 ] \nonumber \\
&+& \frac{1}{2}[ (\nabla b({\vec r}))^2 + m_{\rho}^2 b({\vec r})^2 ] 
+ \frac{1}{2} (\nabla A({\vec r}))^2 , 
\label{qmclag}
\end{eqnarray}
where
$\psi({\vec r})$,$\sigma({\vec r})$, $\omega({\vec r})$, $b({\vec r})$, 
and $A({\vec r})$ are the nucleon, $\sigma$, $\omega$, $\rho$,
and Coulomb fields, respectively.
Note that only the time components of the $\omega$ (a vector-isoscalar meson) 
and the neutral $\rho$ (a vector-isovector meson) are kept 
in the mean field approximation.
These five fields now depend on position $\vec{r}$, 
relative to the center of the nucleus.
The spatial distributions are determined by solving the equations of motion
self-consistently. The key difference between QMC  
and QHD\cite{QHD} lies only in the $\sigma NN$ coupling constant,
 $g_\sigma(\sigma({\vec r}))$ , which
depends on the scalar field in QMC, while it remains  constant in QHD. 
(In practice this is well approximated by
$g_\sigma[1-(a_N/2)g_\sigma\sigma(r)]$.)  
The coupling constants $g_\sigma$, $g_\omega$ and  $g_{\rho}$  
are fixed to reproduce the saturation properties and the bulk symmetry energy
of nuclear matter.
The only free parameter, $m_\sigma$, which controls the range of the 
attractive interaction, and therefore affects the nuclear surface slope 
and its thickness,
is fixed by fitting the experimental rms charge radius of $^{40}Ca$, while
keeping the ratio $g_\sigma/m_\sigma$ fixed, as constrained 
by the properties of nuclear matter.

The quark wave function, as well as the nucleon wave function
(both are Dirac spinors), are determined 
once a solution to equations of motion are found self-consistently.
The orbital electromagnetic form factors for a bound proton, 
in  local density approximation,
are  simply given by
\begin{equation}
G_{E,M}^\alpha(Q^2)= \int G_{E,M} (Q^2,\rho_B(\vec{r})) 
\rho_{p\alpha}(\vec{r}) \,d\vec{r},
\end{equation}
where $\alpha$ denotes a specified orbit with appropriate quantum numbers,
and $G_{E,M}(Q^2,\rho_B(\vec{r}))$ is  the density-dependent form factor 
of a ``proton'' immersed in nuclear matter with local baryon density, 
$\rho_B(\vec{r})$\footnote{In a more sophisticated treatment, 
for example, using a full distorted wave calculation, the weighting may 
emphasize the nuclear surface somewhat more\cite{kelly}.}.
Using the nucleon shell model wave functions, 
the local  baryon density and the local proton density in the specified
 orbit, $\alpha$, are easily evaluated as 
\begin{eqnarray}
\rho_B(\vec{r}) &=&  \sum_\alpha^{\rm occ} d_\alpha
\psi^\dagger_\alpha(\vec{r})\psi_\alpha(\vec{r}),
\nonumber \\ 
\rho_{p\alpha}(\vec{r}) &=& (t_\alpha+{1\over 2})
\psi^\dagger_\alpha(\vec{r})\psi_\alpha(\vec{r}),
\end{eqnarray}
where $d_\alpha= (2j_\alpha+1)$ refers to the degeneracy of nucleons occupying
the orbit $\alpha$ and $t_\alpha$ is the eigenvalue of the isospin operator,
$\tau^N_3/2$.
Notice that the quark wavefunction only depends  on the 
surrounding baryon density. Therefore 
this part of the  calculation of $G_{E,M}(Q^2,\rho_B(\vec{r}))$ is
the same  as in our previous publication for nuclear matter\cite{medium}.  


The notable medium modifications of the quark wavefunction inside the bound 
``nucleon'' in QMC include a reduction of its frequency and an enhancement 
of the lower component of the Dirac spinor.
As in earlier work, the corrections arising from recoil and center of mass 
motion for the bag are made  using the Peierls-Thouless
projection method, combined with  Lorentz contraction of the internal
quark wave function and with the perturbative pion cloud added 
afterwards\cite{ltw}. Note that 
possible off-shell effects\cite{offshell} and 
meson exchange currents\cite{MEC} are ignored in the present approach. 
The resulting nucleon electromagnetic form factors agree with  experiment
quite  well in free space\cite{ltw}. 
Because of  the limitations of the bag model
the form factors  are expected to be most reliable at low momentum 
transfer (say, less than 1 $\mbox{GeV}^2$).
To cut down theoretical uncertainties, we prefer to show the ratios
of the form factors with respect to corresponding free space values.
Throughout this work, we use the renormalized $\pi NN$ coupling constant, 
$f^2_{\pi NN} \simeq 0.0771$\cite{Bugg}. 
The bag radius in free space is taken to be 0.8 fm and the current quark 
mass is 5 MeV in the following  figures.

Fig.~\ref{he4.ps} shows the ratio of the electric and magnetic form factors  
for $^4He$ (which has only one state, $1s_{1/2}$) with respect to the 
free space values .
As expected, both the electric and magnetic rms  radii become slightly larger,
while the magnetic moment of the proton increases by about 7\%.
Fig.~\ref{co16B.ps} shows the ratio  of the electric and magnetic form 
factors for $^{16}O$ with respect to the free space values,
which has  one $s$-state, $1s_{1/2}$, and two $p$-states, 
 $1p_{3/2}$ and $1p_{1/2}$. The momentum dependence of the form factors
for the $s$-orbit nucleon is  more supressed as the inner orbit in $^{16}O$ 
experiences a larger average baryon density
than in $^4He$.
The magnetic moment for the  $s$-orbit nucleon  is similar to that in $^4He$, 
but it is reduced  by $2 - 3$\% in the  $p$-orbit.
Since the difference between two $p$-orbits is rather small, we do not
plot the results for $1p_{1/2}$.
For comparison, we  also show in Fig.~\ref{co16B.ps}
the corresponding ratio of form factors (those curves with triangle symbols)
using a variant of QMC where the bag constant is allowed to decrease 
by 10\%\cite{bagconst}.
It is evident that the effect of a possible reduction in  $B$ is quite large 
and will severely reduce the electromagnetic form factors for a bound nucleon
since the bag radius  is quite sensitive to the value of $B$. 

From the experimental point of view, it is more reliable to show the ratio,
 $G_E/G_M$, since it can be derived directly from the ratio of transverse 
to longitudinal polarization of the outgoing proton,
with minimal systematic errors. 
We find that $G_E/G_M$ runs roughly from 0.41  at $Q^2 = 0 $ to 0.28 and 0.20 
at $Q^2 = 1 \mbox{ GeV}^2$ and $2 \mbox{ GeV}^2$, respectively, for a proton
in the $1s$ orbit in $^4He$ or $^{16}O$.
The  ratio of  $G_E/G_M$ with respect to the corresponding free
space ratio is presented in Fig.~\ref{2ratio.ps}.
The result for the $1s$-orbit  in $^{16}O$ is close to that in  $^4He$ 
and   2\% lower than that for the $p$-orbits in $^{16}O$.
The effect on this ratio of ratios of a reduction in $B$ by the maximum
permitted from other constraints\cite{sick} is quite significant,
especially for larger $Q^2$.

For completeness, we have also calculated the orbital electric and magnetic 
form factors for heavy nuclei such as $^{40}Ca$ and $^{208}Pb$.
The form factors for the proton in selected orbits are shown in Fig.4. 
Because of the larger central baryon density of heavy nuclei, 
the proton electric and magnetic form factors 
in the inner orbits ($1s_{1/2}$, $1p_{3/2}$ and $1p_{1/2}$ orbits) 
suffer much stronger medium modifications
than those in light nuclei.
That is to say, the $Q^2$ dependence is further suppressed, while
the magnetic moments appear to be larger.
Surprisingly, the nucleons in peripheral orbits ($1d_{5/2}$, $2s_{1/2}$, 
and $1d_{3/2}$ for $^{40}Ca$
and $2d_{3/2}$, $1h_{11/2}$, and $3s_{1/2}$ for $^{208}Pb$) 
still show significant medium effects, 
comparable to those in $^4He$.

    Finally, we would like to add some comments on the 
magnetic moment in a nucleus.  In the present calculation, we 
have only calculated the contribution from the intrinsic 
magnetization (or spin) of the nucleon, which is modified 
by the scalar field in a nuclear medium\cite{saito95}.  
As shown in the figures we have found that the intrinsic 
magnetic moment is enhanced in matter because of the change in the quark 
structure of the nucleon.  We know, however, that there 
are several, additional contributions to the nuclear magnetic 
moment, such as  meson exchange currents, higher-order correlations, etc.  
As is well 
known in relativistic nuclear models like QHD, there is a 
so-called magnetic moment problem in mean-field approximation\cite{walecka95}.
To cure this problem, one must 
calculate the convection current matrix element within relativistic 
random phase approximation (RRPA)\cite{kurasawa85}.
However, at high momentum transfer we expect 
that it should be feasible to detect the enhancement of the intrinsic 
spin contribution which we have predicted because the long-range 
correlations, like RRPA, should decrease much faster in that region.

In summary, we have calculated the electric and magnetic form factors 
for the proton, bound in  specific orbits, for several
 closed shell, finite nuclei.
Generally the electromagnetic rms radii and the magnetic moments of the 
bound proton are increased by the medium modifications.
While the difference between the nucleon form factors for 
orbits split by the spin-orbit force is very small,  the difference 
between inner and peripheral orbits is considerable. 
In view of current experimental developments, including the ability to 
precisely measure electron-nucleus quasielastic scattering 
polarization observables, it should be possible to detect differences 
between the form factors in different shell model orbits.
The current and future experiments at
TJNAF and Mainz therefore promise to provide vital information 
with which to guide and constrain dynamic 
microscopic models for finite  nuclei, and perhaps unambiguiously isolate
a signature for the role of quarks.

We would like to acknowledge useful discussions with C. Glashausser and
a helpful communication from J.J. Kelly.
This work was supported by the Australian Research Council.

\begin{figure}
\vspace{2.5cm}
\centering{\
\epsfig{file=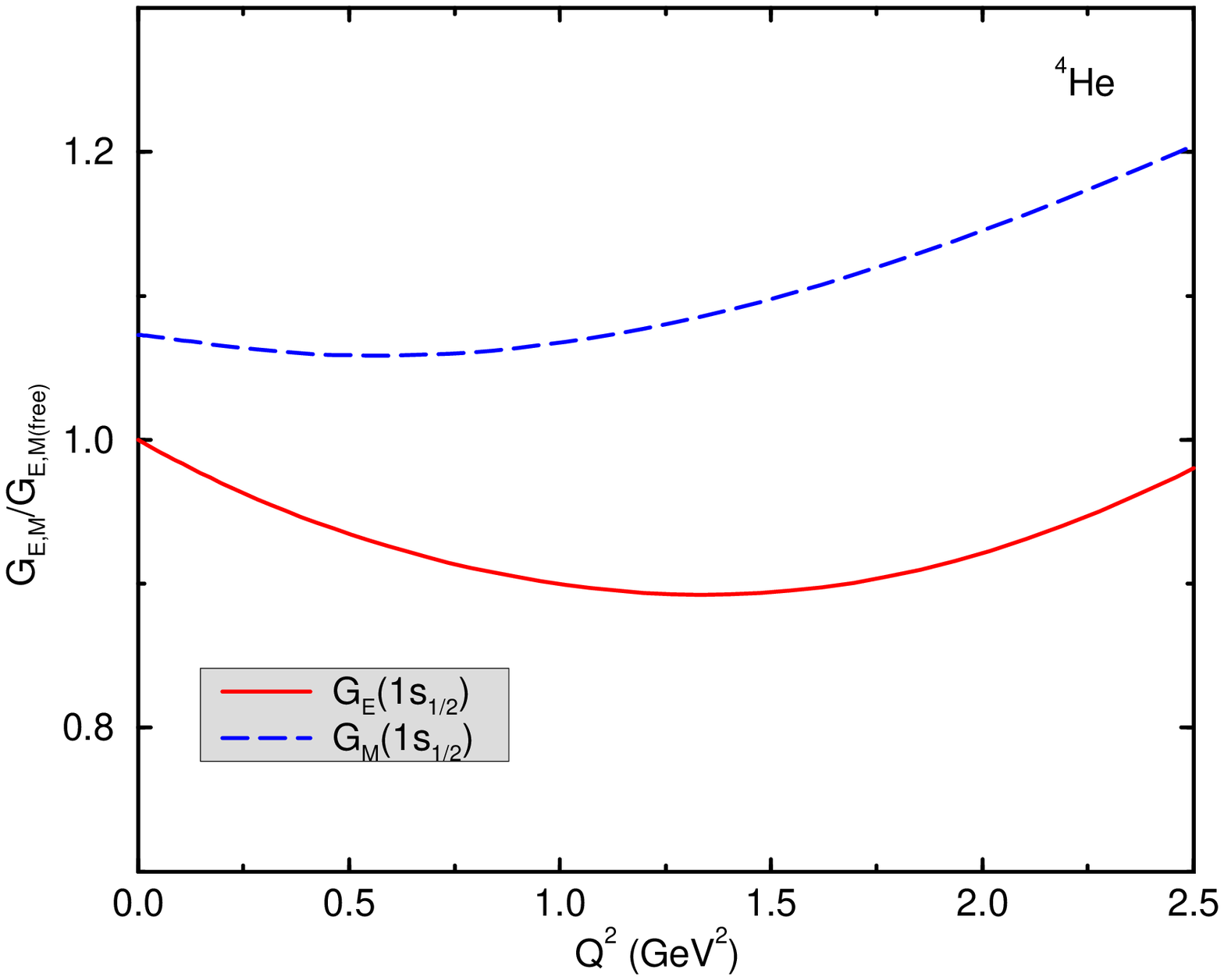,width=12cm}
\caption{Ratio of in-medium to free space electric and magnetic form factors
for the proton in $^4He$.
(The free bag radius was taken to be $R_0=0.8$ fm in all figures.)}
\label{he4.ps}}
\end{figure}

\begin{figure}
\vspace{2.5cm}
\centering{\
\epsfig{file=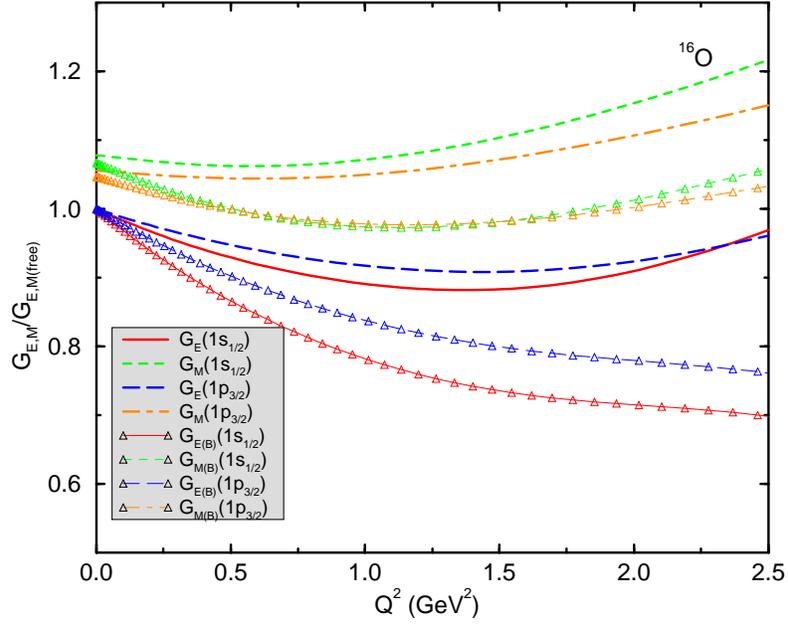,width=12cm}
\caption{Ratio of in-medium to free space electric and magnetic form factors
for the  $s$- and $p$-shells of $^{16}O$.
The curves with triangle symbols represent the corresponding ratio calculated
in a variant of QMC with a 10\% reduction of the bag constant, $B$.}
\label{co16B.ps}}
\end{figure}

\begin{figure}
\vspace{2.5cm}
\centering{\
\epsfig{file=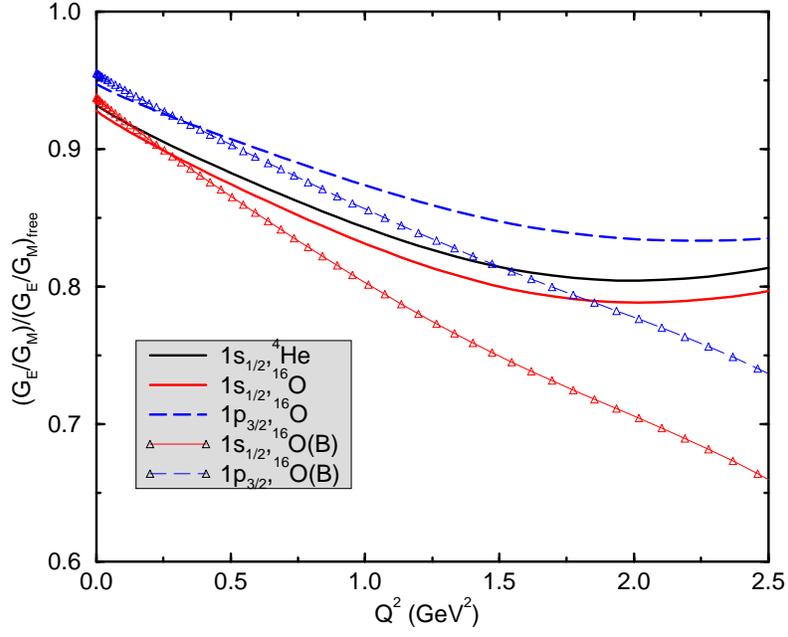,width=12cm}
\caption{Ratio of electric and magnetic form factors in-medium, divided
by the free space ratio. As in previous figure, curves with triangle symbols 
represent the corresponding 
values calculated in a variant of QMC  with a 10\% reduction of $B$.}
\label{2ratio.ps}}
\end{figure}

\begin{figure}
\vspace{2.5cm}
\centering{\
\epsfig{file=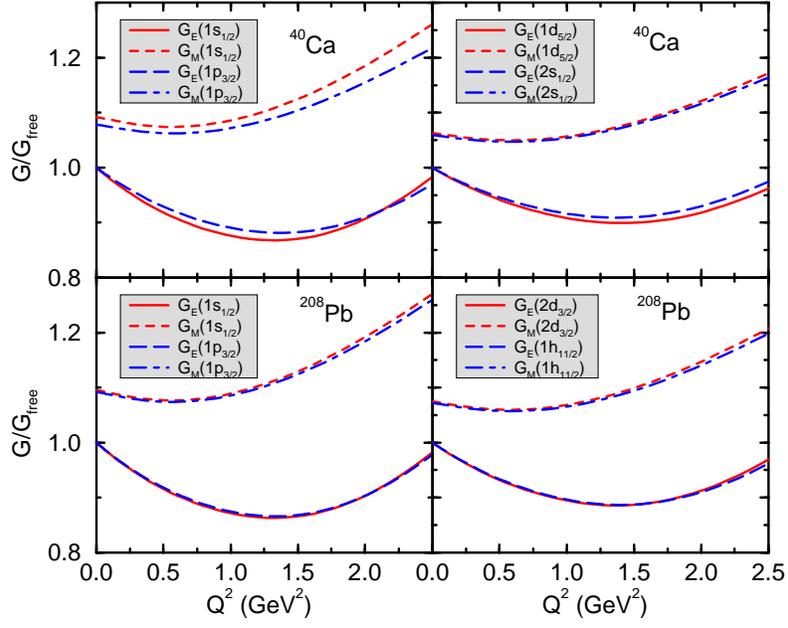,width=12cm}
\caption{Ratio of in-medium to free space electric and magnetic form factors
in specific orbits, for $^{40}Ca$ and $^{208}Pb$.}
\label{ca-pb.ps}}
\end{figure}
\end{document}